\documentclass[runningheads]{llncs}
\usepackage{graphicx}
\usepackage{array}
\usepackage{multirow}
\usepackage{bbding}
\newcolumntype{x}[1]{>{\centering\arraybackslash\hspace{0pt}}p{#1}}
\makeatletter
\newcommand{\printfnsymbol}[1]{%
  \textsuperscript{\@fnsymbol{#1}}%
}
\makeatother

\begin{document}
\title{Fetal Pose Estimation in Volumetric MRI using a 3D Convolution Neural Network}

\author{Junshen Xu\inst{1}\thanks{equal contribution}\inst{(}\Envelope\inst{)} \and
Molin Zhang\inst{2}\printfnsymbol{1} \and
Esra Abaci Turk\inst{3} \and
Larry Zhang\inst{4} \and
\\ Ellen Grant\inst{3,5} \and
Kui Ying\inst{2} \and
Polina Golland\inst{1,4} \and
Elfar Adalsteinsson\inst{1,6}
}
% index{Xu, Junshen}
% index{Zhang, Molin}
% index{Turk, Esra Abaci}
% index{Zhang, Larry}
% index{Ellen, Grant}
% index{Ying, Kui}
% index{Golland, Polina}
% index{Adalsteinsson, Elfar}

%
\authorrunning{J. Xu et al.}
\titlerunning{Fetal Pose Estimation in 3D MRI}
% First names are abbreviated in the running head.
% If there are more than two authors, 'et al.' is used.
%
\institute{Department of Electrical Engineering and Computer Science, MIT, \\ Cambridge, MA, USA \\\email{elfar@mit.edu} \and
Department of Engineering Physics, Tsinghua University, Beijing, China \and
Fetal-Neonatal Neuroimaging and Developmental Science Center, \\ Boston Children’s Hospital, Boston, MA, USA \and
Computer Science and Artificial Intelligence Laboratory, MIT, \\
Cambridge, MA, USA \and
Harvard Medical School, Boston, MA, USA \and
Institute for Medical Engineering and Science, MIT, Cambridge, MA, USA}

\maketitle              
% typeset the header of the contribution
%
\begin{abstract}
The performance and diagnostic utility of magnetic resonance imaging (MRI) in pregnancy is fundamentally constrained by fetal motion. Motion of the fetus, which is unpredictable and rapid on the scale of conventional imaging times, limits the set of viable acquisition techniques to single-shot imaging with severe compromises in signal-to-noise ratio and diagnostic contrast, and frequently results in unacceptable image quality. Surprisingly little is known about the characteristics of fetal motion during MRI and here we propose and demonstrate methods that exploit a growing repository of MRI observations of the gravid abdomen that are acquired at low spatial resolution but relatively high temporal resolution and over long durations (10-30 minutes). We estimate fetal pose per frame in MRI volumes of the pregnant abdomen via deep learning algorithms that detect key fetal landmarks. Evaluation of the proposed method shows that our framework achieves quantitatively an average error of 4.47 mm and 96.4\% accuracy (with error less than 10 mm). Fetal pose estimation in MRI time series yields novel means of quantifying fetal movements in health and disease, and enables the learning of kinematic models that may enhance prospective mitigation of fetal motion artifacts during MRI acquisition. 

\end{abstract}

\section{Introduction}

Estimation of fetal pose from volumetric MRI in pregnancy has applications that include motion tracking and prospective artifact mitigation during diagnostic imaging, retrospective analysis and evaluation of movement by the fetus, as well as the establishment of kinematic models of fetal movement during MRI. Prior work in fetal motion includes methods that rely on simple indices for fetal motion analysis and quantification, such as the angle of the fetal body axes with respect to the maternal body \cite{biglari2016fetal} and maternal perception of fetal movements \cite{heazell2008methods}.

Although pose estimation for the human (adult) body is an established domain in computer vision \cite{newell2016stacked}, to the best of our knowledge, no work has demonstrated fetal pose estimation over time in medical images by MRI. In contrast to human pose estimation from 2D photography, in fetal pose estimation we need to predict 3D pose from dense volumetric data, which increases the computational burden. Further complicating the task is the variable orientation of the fetus within the mother, rapid growth and change in fetal features over gestational age, and poor-quality observations of ground truth pose. 

In pose estimation, handcrafted features such as graphical models and tree-based methods typically suffer from low accuracy and low processing speed while recent developments in deep learning have demonstrated great success in computer vision with acceleration by GPUs and the capability to learn high-level features from data. Consequently, deep convolution neural networks have also found their way into human pose estimation and achieved state-of-the-art results.

\begin{figure}[h]
\centering
  \begin{tabular}{c @{\qquad} c }
    \includegraphics[width=.25\linewidth]{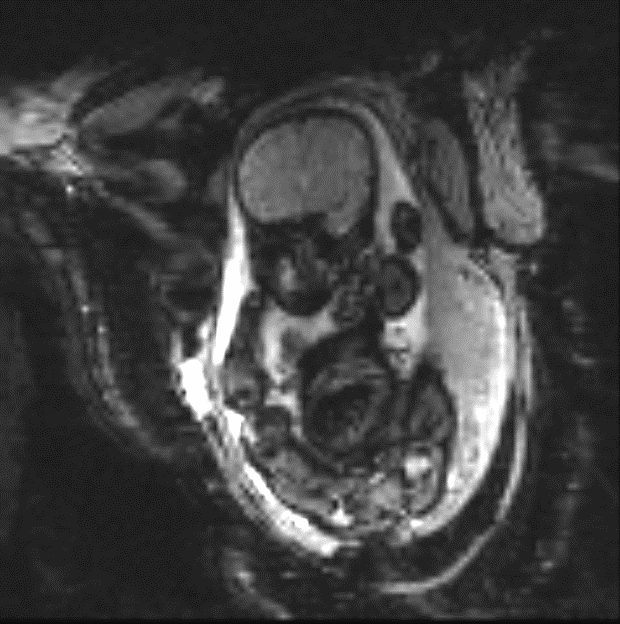} &
    \includegraphics[width=.35\linewidth]{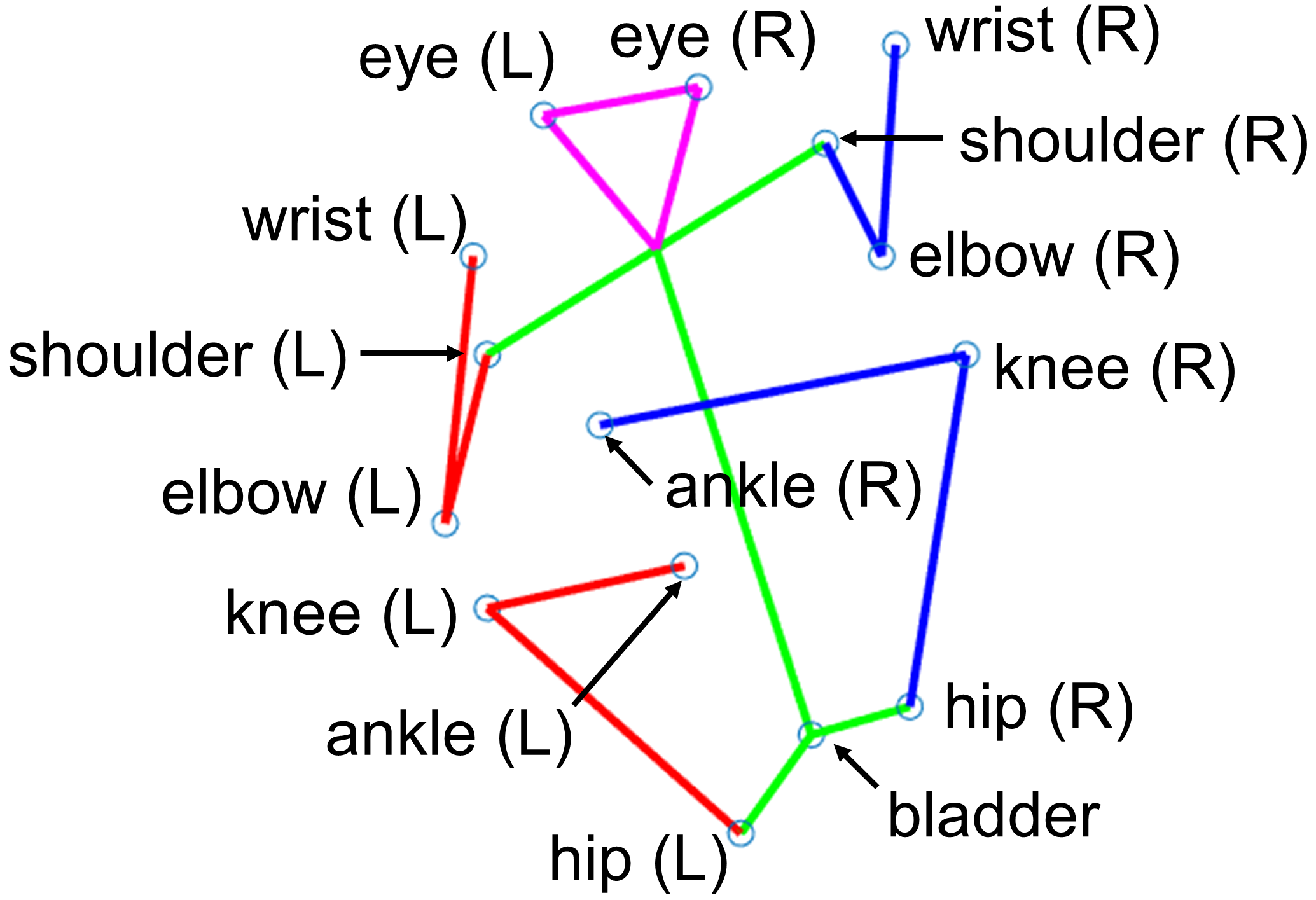} \\
    \small (a) & \small (b)
  \end{tabular}
  \caption{(a) A representative slice from one MRI volume used in this study, and (b) an example of the associated 15 keypoints that characterize fetal pose in three dimensions at a single 3.5-second time frame extracted from a 30-minute observation of the fetus by MRI.}
    \label{fig:keypoints}
\end{figure}

In an ongoing study of placental function by EPI BOLD imaging time series (see Figure \ref{fig:keypoints} (a)), we have built an archive of over 70 subjects, each with 200-500 time frames of EPI volumes, imaged continuously over 10-30 minute observation intervals and resulting in over 18,000 EPI volumes. By visual inspection, the fetal pose can be inferred from these data but manual labeling of keypoints for pose estimation (see Figure \ref{fig:keypoints} (b)) across these volumes is prohibitive and here we propose a method based on deep neural networks to identify fetal key points.

%, which exploits texture information at various scales and the correlation among features. 

We propose, demonstrate, and characterize the performance of a two-stage framework for fetal pose estimation in 3D MRI using deep learning, where we first generate heatmaps for each fetal keypoint using a convolution network and then infer fetal pose from heat maps using a Markov Random Field (MRF) that exploit anatomically rational information about connections between keypoints. Evaluation of performance shows that the proposed method achieves a mean error of 4.47 mm and a percentage of correct detection of 96.4\%. Further, computation time of our pipeline is less than 1 s/volume, which potentially enables low-latency tracking of fetal pose during diagnostic MRI in pregnancy.

%Another problem related to fetal pose estimation is landmark localization in medical images\cite{}, since the pose is usually represented by a %set of keypoints. Our task differs from the standard problem in several aspects. First, landmark localization usually focuses on an organ or %region, e.g., brain, heart and knee, while, in our case, the pose of the whole fetal body is our target. Second, keypoints in pose estimation %have more connection and constraints. For example, the distance between two joints is constant if they are connected by a bone. Such %information can make a difference in the performance of pose estimation. Third, the orientation of object of interest in landmark %localization, e.g., brain, is usually unchanged. However, in fetal imaging, the fetal orientation varies greatly, which makes the problem more %complicated.

%Briefly state the architecture? Summarize performance, runtime, etc? 

\section{Methods}

\subsection{Pose Estimation Framework}

\begin{figure}[h]
\centering
\includegraphics[width=0.7\textwidth]{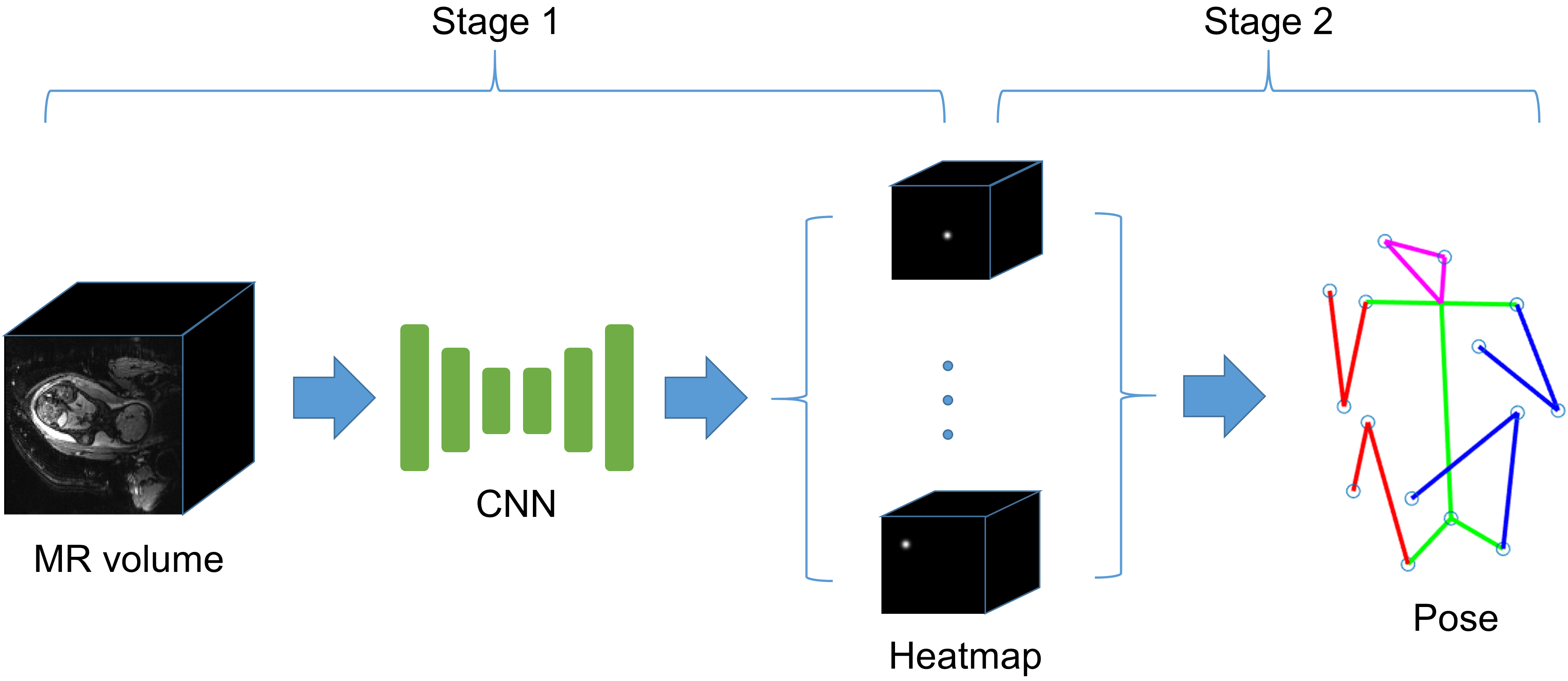}
\caption{The framework of fetal pose estimation in 3D MRI which consists of two stages. Stage 1: generate 3D heatmaps of each keypoint from the input MR volume. Stage 2: estimate keypoint locations from heatmaps.} \label{fig:framework}
\end{figure}

Exploring the idea of heatmap prediction in human pose estimation \cite{newell2016stacked}, here we propose a two-stage framework for fetal pose estimation in 3D MRI using deep learning (see Fig. \ref{fig:framework}). In the first stage, a CNN is used to generate heatmaps from input MR volume, which produce per-pixel likelihoods for keypoints on the fetal skeleton. However, the generated heatmap may have multiple local maxima and simply using max activating location as prediction may lead to low accuracy.

To address this problem, a second stage is proposed to infer location from estimated heatmaps, exploiting the constraints of fetal pose to refine the results. We model the fetal pose as a MRF, where each keypoint of fetus is represented by a node in the graph and the states are the plausible locations of the keypoint. The final prediction is generated by performing inference on this MRF.

The following subsections describe the proposed framework in detail.

\subsection{Heatmap Prediction using CNN:}

\begin{figure}[h]
\centering
\includegraphics[width=0.8\textwidth]{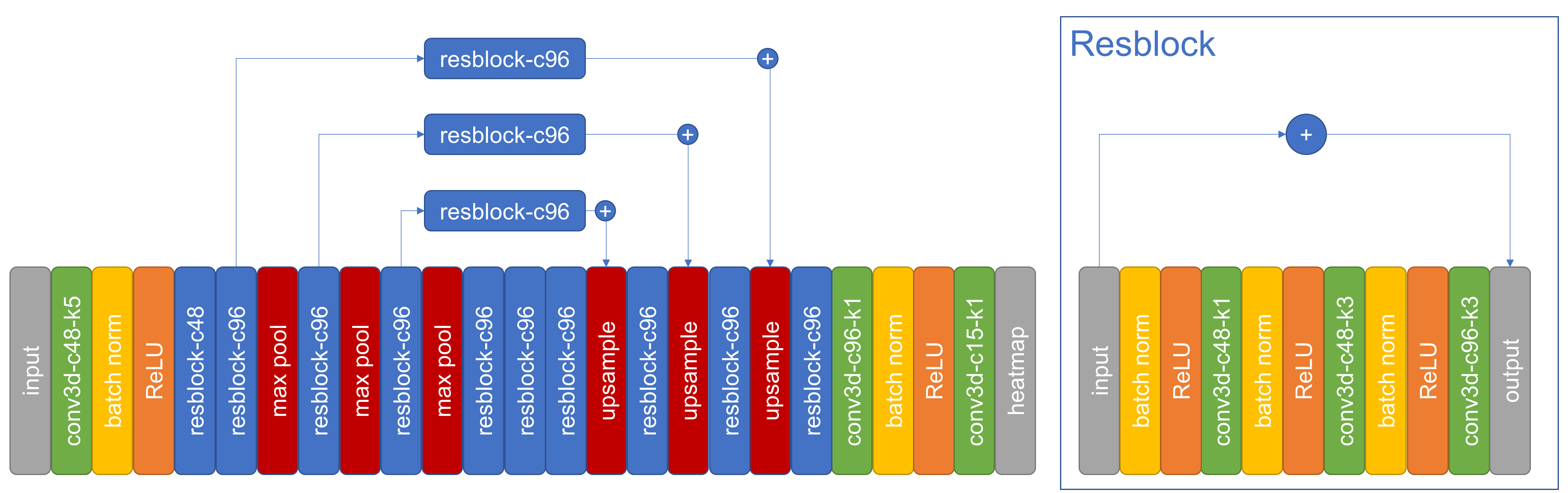}
\caption{Left: architecture of 3D hourglass network for heatmap prediction. Right: structure of resblock.} \label{fig:network}
\end{figure}

Inspired by the successful application of hourglass networks in human pose estimation \cite{newell2016stacked}, we propose a 3D hourglass network for heatmap prediction of fetal keypoints. The overall architecture of the proposed network is shown in Fig. \ref{fig:network}. The network is based on the encoder-decoder structure which is motivated by the idea of capturing multi-scale information. In pose estimation, while local evidence, e.g., local contrast, is important for identification of keypoint, global information can help resolve ambiguity, such as fetus' orientation and relative position of other joints or body parts. In each scale of the network, resblocks with 3D convolution layers are used to extract features. To recover loss of high resolution information in downscale-upscale structure, skipped connections with element-wise addition are adopted to connect symmetric scales.

The CNN tries to learn a mapping from MR images to target heatmaps, which is generated by placing a Gaussian distribution with $\sigma=2$ on the ground-truth position and stacking together. So the output heatmaps will be of the same spatial dimensions but have $J$ channels, where $J$ is the number of keypoints need to predict. The loss function used for training is the mean-squared error (MSE) between the predicted heatmap and target heatmap. Instead of using the whole volume, 3D patches with size of $64\times64\times64$ are used as input for training. This strategy can reduce GPU memory usage, enabling mini-batch training. Since the network is fully convolutional, in inference, the whole 3D MR volumes are fed into the network to generate heatmap of full scale. 

\subsection{Location Estimation from Heatmap:} Given the output heatmap from CNN, the second stage of the pose estimation framework is to estimate location of each keypoint. Let $x_i$ and $H_i$ be the location and heatmap of the $i-$th keypoint, $i=1,...,J$. Let $x=(x_1,...,x_J)$. Then one simple idea to infer keypoint positions from heatmaps is taking the max activating location of each heatmap%, i.e.,
%\begin{equation}
%    \hat{x}_i=\arg\max_{x_i} H_i(x_i)
%\end{equation}
However, this method handles each keypoint independently and does not make use of the connection between keypoints, e.g., the distance between two joints should be a constant if they are connected by bones. To exploiting these connections, we model the fetal pose as a MRF, where each keypoint correspond to a node in the graph and connections of keypoints are represented as edges in the graph. The states $\mathcal{S}_i=\{x_i^{(1)}, ..., x_i^{(L)}\}$ for node $i$ is the top-$L$ local maxima in heatmap $i$. Our prediction of fetal pose would be a particular configuration of the MRF, i.e., $\hat{x}\in\mathcal{S}_1\times\cdots\times\mathcal{S}_J$. Each configuration is assigned an energy, $E(x),$ defined as
\begin{equation}
     E(x)= \sum_{i=1}^J \varphi_i(x_i) + \sum_{(i,j)\in B}\phi_{i,j}(x_i, x_j)
     \label{eqn:energy}
 \end{equation}
 where $B$ is the set of connections. A low energy of a configuration implies high probability. Therefore, the inference is equivalent to finding the configuration with lowest energy%:
%\begin{equation}
%     \hat{x}=\arg\min_x E(x)
%     \label{eqn:opt}
%\end{equation}

Since the heatmap can be considered as a surrogate for the probability distribution of the corresponding keypoint, the unary term in energy function $F$ can be modeled as
\begin{equation}
    \varphi_i(x_i)=-\log H_i(x_i)
\end{equation}
As for the pairwise term, we define $\phi_{i,j}$ as a quadratic function of $||x_i-x_j||_2$, the distance between keypoint $i$ and $j$.
\begin{equation}
    \phi_{i,j}(x_i, x_j)=-\frac{\alpha(||x_i-x_j||_2/r_t-\mu_{ij})^2}{\sigma_{ij}^2},
\end{equation}
where $r_t$ is the mean bone length at gestational age $t$, so that $||x_i-x_j||_2/r_t$ can be regarded as the distance of two keypoints normalized by gestational age. $\mu_{ij}$ and $\sigma_{ij}^2$ are the mean and variance of the normalized distance, which are estimated from training data. $\alpha$ is the regularization weight. The optimization problem is solved by a belief propagation algorithm \cite{schmidt2012ugm}. 

\section{Experiments and Results}

\subsection{Dataset}

The data for this study consist of volumetric MRI time series from imaging of 70 mothers pregnant with singletons at a gestational age ranging from 25 to 35 weeks. MRIs were acquired on a 3T Skyra scanner (Siemens Healthcare, Erlangen, Germany). Multislice, single-shot, gradient echo EPI sequence was used for acquisitions with in-plane resolution of $3\times3$mm$^2$, slice thickness of 3 mm, mean matrix size = $120\times120\times80$; TR=$5-8$s, TE=$32-38$ms, FA=90$^{\circ}$. Each subject was scanned for 10 to 30 min.

Similar to the task of adult human pose estimation, we model the pose of a fetus with a set of keypoints. We chose fifteen keypoints (ankles, knees, hips, bladder, shoulders, elbows, wrists and eyes) to capture pose and labeled manually, with a representative example shown in Fig. \ref{fig:keypoints}(b). These fifteen landmarks were selected as keypoints as they capture gross fetal anatomy that is critical in subsequent motion analysis, and they presented with adequate image contrast to be relatively robustly observed in the MR volumes, thus mitigating the error and noise in labelling. In total, 1705 MR volumes were labelled, 1028($\sim60\%$) for training, 240($\sim15\%$) for validation and 437($\sim25\%$) for testing, where the testing set consists of subjects different from training and validation sets.

In order to improve the generalization capacity and avoid overfitting, several data augmentation techniques were used, including intensity scaling, 3D rotation and flipping.
% (1) Intensity Scaling: To increase the network's robustness to difference in MR contrast,  we randomly scale each volume's intensity values with a factor uniformly sampled from 0.9 to 1.1. (2) Rotation: We randomly rotate the input MR volume along different axes by $90^{\circ}$ so that the trained network is robust to the change of fetal orientations due to motion. (3) Flipping: We also augment the data by flipping the volumes along different axes, in order to make the network aware of global features that help distinguish between left- and right-sided keypoints.

\subsection{Experiments Setup}

All experiments were performed on a server with an Intel Xeon E5-1650 CPU, 128GB RAM and a NVIDIA TITAN X GPU. Neural networks were implemented with TensorFlow and for optimization we use Adam with an initial learning rate of $5\times10^{-3}$, weight decay of $1\times10^{-4}$ and the restart strategy \cite{loshchilov2017fixing}. The networks are trained for 200 epochs. For the second stage, we set $L=3$ and $\alpha=1$.

\subsection{Results}

In this section, we evaluate the proposed pipeline for fetal pose estimation. First, we evaluate the proposed 3D hourglass network (HG) with max activating location of the heatmap as final prediction. For comparison, 3D UNet\cite{cciccek20163d} is used in our experiment, which has been used for heatmap regression\cite{payer2016regressing}. Finally, we examine the whole pipeline by combine the CNN-based heatmap regression and MRF. These models are denoted as UNet-M and HG-M respectively.

Several metrics are used for evaluation: a) Percentage of Correct Keypoint (PCK), where a detected keypoint is considered correct if the distance between the predicted and the true keypoint is within a certain threshold, b) mean error (in mm),i.e., the mean distance between the predicted and the ground-truth keypoint, and c) median of error.

\begin{figure}[h]
\centering
\includegraphics[width=0.8\textwidth]{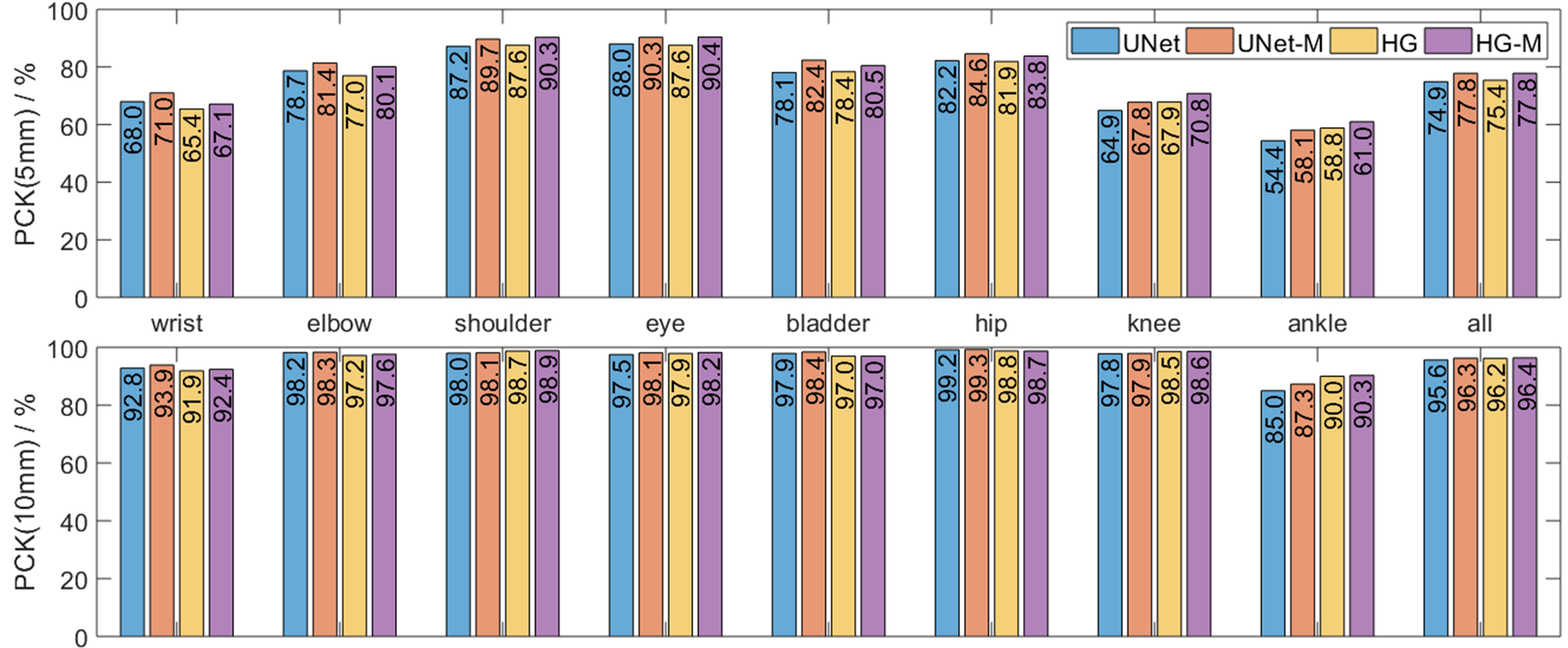}
\caption{PCK with two threshold, 5mm (1.67 pixel) and 10mm (3.33 pixel) for different keypoints.} \label{fig:pck}
\end{figure}

\begin{table}[h]
\caption{mean and median of error of different models.}
\label{tab1}
\begin{center}
\begin{tabular}{x{1.3cm}|x{1.2cm}|x{0.92cm}x{1cm}x{1.2cm}x{0.92cm}x{1.2cm}x{0.92cm}x{0.92cm}x{0.92cm}x{0.92cm}}
\hline
metric &method & wrist & elbow & shoulder & eye & bladder & hip & knee & ankle & all\\
\hline
% \multirow{4}{*}{\shortstack{PCK\\(5mm)}}
% &UNet  &68.0 &78.7 &87.2 &88.0 &78.1 &82.2 &64.9 &54.4 &74.9 \\
% &UNet-M&\textbf{71.0} &\textbf{81.4} &89.7 &90.3 &\textbf{82.4} &\textbf{84.6} &67.8 &58.1 &\textbf{77.8} \\
% &HG    &65.4 &77.0 &87.6 &87.6 &78.4 &81.9 &67.9 &58.8 &75.4 \\
% &HG-M  &67.1 &80.1 &\textbf{90.3} &\textbf{90.4} &80.5 &83.8 &\textbf{70.8} &\textbf{61.0} &\textbf{77.8}  \\
% \hline
% \multirow{4}{*}{\shortstack{PCK\\(10mm)}}
% &UNet  &92.8      &98.2      &98.0         &97.5    &97.9        &99.2    &97.8     &85.0      &95.6  \\
% &UNet-M&\textbf{93.9}&\textbf{98.3}&98.1 &98.1 &\textbf{98.4}&\textbf{99.3}    &97.9     &87.3      &96.3  \\
% &HG    &91.9      &97.2      &98.7  &97.9    & 97.0     &98.8    &98.5     & 90.0     &96.2  \\
% &HG-M  &92.4      &97.6      &\textbf{98.9}   &\textbf{98.2}    &97.0      &98.7    &\textbf{98.6}     &\textbf{90.3}      &\textbf{96.4}\\
% \hline
\multirow{4}{*}{\shortstack{median\\(mm)}}
&UNet  &3.84              &3.43               &2.87            &2.74             &3.20          &\textbf{3.12 }   &4.00     &4.42      &3.47  \\
&UNet-M&3.84              &3.43               &2.87            &2.73             &\textbf{3.19} &\textbf{3.12}    &3.99    &4.36      &3.46  \\
&HG    &\textbf{3.82}     &3.42               &\textbf{2.83}   &\textbf{2.72}    &3.37          &3.16        &3.87     &\textbf{4.15}     & \textbf{3.42} \\
&HG-M  &\textbf{3.82}     &\textbf{3.41}      &\textbf{2.83 }  &\textbf{2.72}    &3.36          &3.16         &\textbf{3.86}     &\textbf{4.15}     &\textbf{3.42}\\
\hline
\multirow{4}{*}{\shortstack{mean\\(mm)}}
&UNet  &7.34&4.06&4.27&3.96&4.48&3.33&5.19&10.2&5.41\\
&UNet-M&\textbf{5.64}&\textbf{3.81}&3.75&3.29&\textbf{3.52}&\textbf{3.23}&4.84&8.18&4.60\\
&HG    &7.48& 4.81& 3.24& 3.35& 4.69& 3.58& 4.39& 7.49&4.89\\
&HG-M  &6.37& 4.11& \textbf{3.10}& \textbf{3.28}& 4.12& 3.33&\textbf{4.19}& \textbf{7.07}& \textbf{4.47}\\
% \multirow{8}{*}{\shortstack{mean\\error\\(std)}}
% &\multirow{2}{*}{UNet}
% &7.34&4.06&4.27&3.96&4.48&3.33&5.19&10.2&5.41\\
% &&(19.7)&(6.80)&(10.3)&(11.8)&(12.3)&(2.70)&(8.87)&(20.3)&(11.6)\\
% &\multirow{2}{*}{UNet-J}
% &\textbf{5.64}&\textbf{3.81}&3.75&3.29&\textbf{3.52}&\textbf{3.23}&4.84&8.18&4.60\\
% &&(\textbf{11.8})&(\textbf{4.47})&(7.53)&(5.12)&(\textbf{2.57})&(\textbf{2.33})&(7.18)&(15.7)&(7.40)\\
% &\multirow{2}{*}{HG}
% &7.48& 4.81& 3.24& 3.35& 4.69& 3.58& 4.39& 7.49&4.89\\
% &&(19.1)&(11.0)&(4.99)&(4.81)&(11.8)&(6.28)&(4.20)&(14.9)&(9.49)\\
% &\multirow{2}{*}{HG-J}
% &6.37& 4.11& \textbf{3.10}& \textbf{3.28}& 4.12& 3.33&\textbf{4.19}& \textbf{7.07}& \textbf{4.47}\\
% &&(12.5)&(6.64)&(\textbf{4.06})&(\textbf{4.69})&(7.70)&(2.62)&(\textbf{2.76})&(\textbf{13.0})&(\textbf{6.69})\\
\hline
\end{tabular}
\end{center}
\end{table}

\begin{table}[h]
\caption{Computation time and number of parameters of different networks.}
\label{tab2}
\begin{center}
\begin{tabular}{x{2cm}x{5cm}x{4cm}}
\hline
network & computation time (ms/volume) & number of parameters\\
\hline
UNet & 271 & 22M\\
HG   & 225 & 3.5M\\
\hline
\end{tabular}
\end{center}
\end{table}

\begin{figure}[h]
\centering
  \begin{tabular}{c @{ } c }
    \includegraphics[width=.2\linewidth]{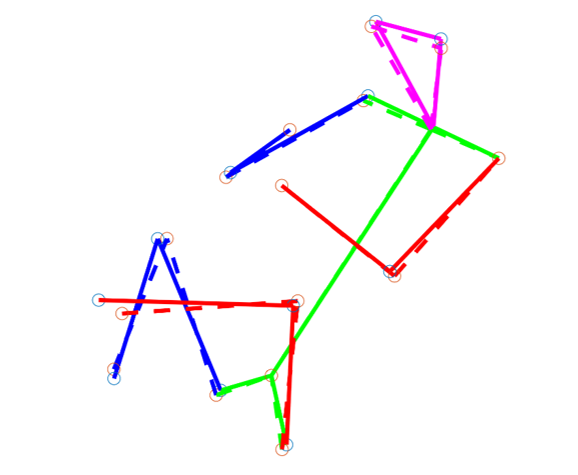} &
    \includegraphics[width=.4\linewidth]{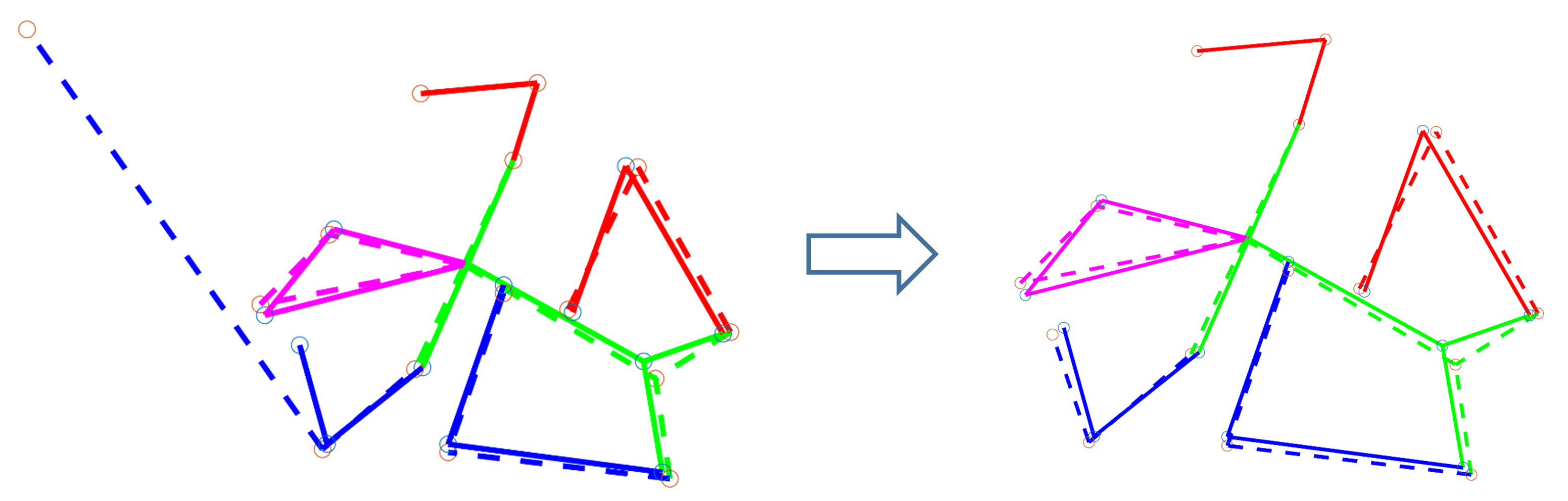} \\
    \small (a) & \small (b)
  \end{tabular}
  \caption{(a) An example of fetal pose successfully predicted by the max activating location of heatmaps, where solid lines are the ground-truth pose and dashed lines are the predicted pose. (b) A failed case of fetal pose estimation with max activation (left), and the corresponding successful result after processed by MRF (right).}
    \label{fig:pose}
\end{figure}

Fig \ref{fig:pck} shows PCK with two threshold, 5mm (1.67 pixel) and 10mm (3.33 pixel) while the mean and median of error of different models are illustared in table \ref{tab1}. Applying the proposed pipeline, 96.4\% of the keypoints are located correctly (with error $<$ 10mm) and the mean distance between predicted and ground-truth keypoints is 4.47mm (1.5 pixel). Besides, we see that, in average, the proposed 3D hourglass network has similar performance compared to 3D UNet. However, as illustrated in table \ref{tab2}, the number of parameters of UNet is 6 times as large as that of hourglass network, indicating that the proposed network is more compact and efficient. The main reason is that the hourglass network use elementwise sum instead of concatenate in skip connection and fix the number of channels across different scales. We also notice that the second stage Markov network refinement improves the performance upon CNN heatmap regression, in terms of PCK as well as mean error. As illustrated in Fig. \ref{fig:pose}(b), fetal pose estimation based on max activating location of heatmap may result in irrational prediction. Such error is corrected in the MRF refinement by making a trade-off between prior information of keypoint connections and heatmaps generated by the CNN. As for computation time, the proposed 3D hourglass network runs at a speed of 225 ms/volume on a GPU and solving the optimization problem for inferring keypoint locations from heatmaps takes 290 ms/volume on CPU. Therefore, the end-to-end processing time of the whole pipeline is less than 1 s/volume and therefore shorter than the temporal resolution in the current fetal MR protocol, which potentially enables low latency tracking of fetal pose in fetal MR imaging.

\section{Conclusions}

In this work, we proposed a two-stage deep learning framework for fetal pose estimation in 3D MRI. The proposed method achieves mean error of 4.47 mm ($\sim$ 1.5 pixels) and percentage of correct detection of 96.4\%, which indicates that deep neural networks are able to identify key features for fetal pose estimation from time frames in low-resolution, volumetric EPI data from pregnant mothers. Further, the total processing time of the proposed framework is less than 1 s, potentially enabling low latency tracking of fetal pose in fetal MR imaging. Limitations of the current method include a pipeline that was only trained on singleton pregnancies. Also, the current pose detection was performed on each time frame in isolation without utilizing any form of temporal correlations in the MR series. In future work the proposed framework could be extended to work with multiplet pregnancies as well as exploit temporal correlations across volumes in a time sequence. 

Overall, the proposed pipeline could be deployed for fetal motion estimation during MR scanning of pregnant mothers with applications to fetal health and disease, establishment of fetal kinetic motion models, and prospective motion correction with slice-prescription updates for more robust diagnostic fetal and maternal MRI. 

\section*{Acknowledgements}
This research was supported by NIH U01HD087211, NIH R01EB01733, NIH NIBIB NAC P41EB015902 and NIH NICHD U01HD087211.

% \begin{table}
% \caption{Table captions should be placed above the
% tables.}\label{tab1}
% \begin{tabular}{|l|l|l|}
% \hline
% Heading level &  Example & Font size and style\\
% \hline
% Title (centered) &  {\Large\bfseries Lecture Notes} & 14 point, bold\\
% 1st-level heading &  {\large\bfseries 1 Introduction} & 12 point, bold\\
% 2nd-level heading & {\bfseries 2.1 Printing Area} & 10 point, bold\\
% 3rd-level heading & {\bfseries Run-in Heading in Bold.} Text follows & 10 point, bold\\
% 4th-level heading & {\itshape Lowest Level Heading.} Text follows & 10 point, italic\\
% \hline
% \end{tabular}
% \end{table}

% ---- Bibliography ----

% BibTeX users should specify bibliography style 'splncs04'.
% References will then be sorted and formatted in the correct style.

\bibliographystyle{splncs04us}
\bibliography{reference}

\begin{thebibliography}{1}
\providecommand{\url}[1]{\texttt{#1}}
\providecommand{\urlprefix}{URL }
\providecommand{\doi}[1]{https://doi.org/#1}

\bibitem{biglari2016fetal}
Biglari, H., Sameni, R.: Fetal motion estimation from noninvasive cardiac
  signal recordings. Physiological measurement  \textbf{37}(11), ~2003 (2016)

\bibitem{heazell2008methods}
Heazell, A.P., Fr{\o}en, J.: Methods of fetal movement counting and the
  detection of fetal compromise. Journal of Obstetrics and Gynaecology
  \textbf{28}(2),  147--154 (2008)

\bibitem{newell2016stacked}
Newell, A., Yang, K., Deng, J.: Stacked hourglass networks for human pose
  estimation. In: European Conference on Computer Vision. pp. 483--499.
  Springer (2016)

\bibitem{schmidt2012ugm}
Schmidt, M.: Ugm: Matlab code for undirected graphical models. URL http://www.
  di. ens. fr/mschmidt/Software/UGM. html  (2012)

\bibitem{loshchilov2017fixing}
Loshchilov, I., Hutter, F.: Fixing weight decay regularization in adam. arXiv
  preprint arXiv:1711.05101  (2017)

\bibitem{cciccek20163d}
{\c{C}}i{\c{c}}ek, {\"O}., Abdulkadir, A., Lienkamp, S.S., Brox, T.,
  Ronneberger, O.: 3d u-net: learning dense volumetric segmentation from sparse
  annotation. In: International conference on medical image computing and
  computer-assisted intervention. pp. 424--432. Springer (2016)

\bibitem{payer2016regressing}
Payer, C., {\v{S}}tern, D., Bischof, H., Urschler, M.: Regressing heatmaps for
  multiple landmark localization using cnns. In: International Conference on
  Medical Image Computing and Computer-Assisted Intervention. pp. 230--238.
  Springer (2016)

\end{thebibliography}

% \begin{thebibliography}{8}
% \bibitem{ref_article1}
% Author, F.: Article title. Journal \textbf{2}(5), 99--110 (2016)

% \bibitem{ref_lncs1}
% Author, F., Author, S.: Title of a proceedings paper. In: Editor,
% F., Editor, S. (eds.) CONFERENCE 2016, LNCS, vol. 9999, pp. 1--13.
% Springer, Heidelberg (2016). \doi{10.10007/1234567890}

% \bibitem{ref_book1}
% Author, F., Author, S., Author, T.: Book title. 2nd edn. Publisher,
% Location (1999)

% \bibitem{ref_proc1}
% Author, A.-B.: Contribution title. In: 9th International Proceedings
% on Proceedings, pp. 1--2. Publisher, Location (2010)

% \bibitem{ref_url1}
% LNCS Homepage, \url{http://www.springer.com/lncs}. Last accessed 4
% Oct 2017
% \end{thebibliography}
\end{document}